\documentclass[aps,prl,reprint,showpacs,amsmath,amssymb,superscriptaddress,twocolumn]{revtex4-1}

\usepackage{graphicx}
\usepackage{amssymb}
\usepackage{amsmath}
\usepackage{color}

\begin{document}

\title{Coarsening scenarios in unstable crystal growth}

\author{Sofia Biagi}
\email{sofia.biagi@ujf-grenoble.fr}
\affiliation{Dipartimento di Fisica e Astronomia,
Universit\`a di Firenze, via G. Sansone 1, 50019 Sesto Fiorentino, Italy}
\affiliation{Istituto dei Sistemi Complessi, Consiglio Nazionale delle Ricerche, Via Madonna del Piano 10, 50019 Sesto Fiorentino, Italy}
\affiliation{Universit\'e Grenoble 1/CNRS, LIPhy UMR 5588, Grenoble, F-38401, France}
\author{Chaouqi Misbah}
\email{chaouqi.misbah@ujf-grenoble.fr}
\affiliation{Istituto dei Sistemi Complessi, Consiglio Nazionale delle Ricerche, Via Madonna del Piano 10, 50019 Sesto Fiorentino, Italy}
\affiliation{Universit\'e Grenoble 1/CNRS, LIPhy UMR 5588, Grenoble, F-38401, France}
\author{Paolo Politi}
\email{paolo.politi@isc.cnr.it}
\affiliation{Istituto dei Sistemi Complessi, Consiglio Nazionale delle Ricerche, Via Madonna del Piano 10, 50019 Sesto Fiorentino, Italy}

\date{\today}

% Proposed Pacs:
% 68.55.-a Thin film structure and morphology
% 05.70.Ln Nonequilibrium and irreversible thermodynamics
% 81.10.Aj Theory and models of crystal growth

% Other possible Pacs
% 05.45.-a Nonlinear dynamics and nonlinear dynamical systems
% 64.60.Ht Dynamic critical phenomena
% 68.43.Jk Diffusion of adsorbates, kinetics of coarsening and aggregation

\newcommand{\be}{\begin{equation}}
\newcommand{\ee}{\end{equation}}
\newcommand{\bea}{\begin{eqnarray}}
\newcommand{\eea}{\end{eqnarray}}
\newcommand{\bse}{\begin{subequations}}
\newcommand{\ese}{\end{subequations}}
\newcommand{\peq}{p_{\hbox{\tiny eq}}}
\newcommand{\bx}{\mathbf{x}}
\newcommand{\bj}{\mathbf{j}}
\renewcommand{\bm}{\mathbf{m}}
\renewcommand{\l}{\ell}
\renewcommand{\th}{\tilde h}

\newcommand{\comment}[1]{}
\newcommand{\red}[1]{\textcolor{red}{#1}}
\newcommand{\blue}[1]{\textcolor{blue}{#1}}

\begin{abstract}
Crystal surfaces may undergo thermodynamical as well kinetic, out-of-equilibrium instabilities.
We consider the case of mound and pyramid formation, a common phenomenon in crystal growth
and a long-standing problem in the field of pattern formation and coarsening dynamics.
We are finally able to attack the problem analitycally and get rigorous results.
Three dynamical scenarios are possible: perpetual coarsening, interrupted coarsening,
and no coarsening. In the perpetual coarsening scenario, mound size increases in time
as $L\sim t^n$, where the coasening exponent is $n=1/3$ when faceting occurs, otherwise $n=1/4$.
\end{abstract}

\maketitle

%--------------------------------------

{\it Introduction}.---The still ongoing large interest towards instabilities in crystal
growth~\cite{RMP_Chaouqi} can be motivated by its lying in between the practical problem of
controlling the growth processes and the fundamental domain of out-of-equilibrium transitions
and pattern formation. Even if some phenomena may be studied in one spatial dimension \cite{1d},
most interesting physics of surface growth occurs in two spatial dimensions,
where rigorous approaches are rare and numerical simulations are more problematic and
much more time consuming.

The unstable growth of a crystal surface has some similarities
with phase ordering processes, for which the analysis of Bray~\cite{Bray} is the most 
comprehensive and successful theory (see also Refs.~\cite{Cugliandolo,2d} for more recent results). 
However, in two dimensions growth has peculiar features~\cite{review_Paolo}, because
the natural order parameter, the surface slope $\bm$, obeys a special constraint
$(\nabla \times \bm=0$), so that resulting domain walls must be straight, which is not the
case with other phase ordering phenomena.
As a consequence of that, rigorous analytical approaches to crystal growth instabilities
are very rare. A paper by Watson and Norris~\cite{Watson}, based on the principle of maximal
dissipation and a couple of papers~\cite{Kohn,BoLi} providing exact inequalities for the coarsening
exponent are some of the few examples.
Other interesting, albeit heuristic, approaches had contributed significantly to
our understanding of coarsening~\cite{Rost_Krug,Golubovic1997,tang}.

At the general level, we think that mound formation in two dimensional crystal growth,
in spite of being quite an old problem~\cite{Villain}, still lacks a systematic
and rigorous approach.
\comment{In this manuscript we propose to fill this gap, starting from
a well established and accepted continuum description of the growing surface and applying to it
a multiscale analysis which is asymptotically exact and allows to face
at once a large class of models and symmetries. We start by briefly reminding the basic phenomenology
of unstable crystal growth which is of interest to us, then we present an important class of
differential equations we are going to study and the central part of the paper follows: the application of the multiscale analysis and its results.}
In this manuscript we propose to fill this gap. We start by briefly reminding the basic phenomenology of unstable crystal growth which is of interest to us and presenting a well established continuum description of the growing surface. Then the central part of the paper follows:
the application of the phase diffusion concept that allows to face
at once a large class of models and symmetries.
\comment{This method will lead us to phase equations describing the evolution of the stationary periodic states after perturbations of their length scale $\lambda=2\pi/q$ and}
In this approach, the pattern instability is signaled by a negative phase diffusion coefficient, which implies coarsening,
i.e. an increase of the size $L$ of the pattern.
% lasts forever or it interrupts.
\comment{I have changed:
This method is based on a suitable perturbation of a
periodic pattern, whose phase acquires slow time and space dependences. Dynamical evolution
of phases will be described by two coupled diffusion equations, whose coefficients determine
the stability of the periodic pattern, i.e. if the systems coarsens or not and if coarsening
lasts forever or interrupts.}
In the case of perpetual coarsening, the dependence of
the diffusion coefficient on $\lambda$ allows to determine the growth law, $L(t)$.
Our work gives the first rigorous, general framework to study unstable crystal growth and
we provide important results concerning coarsening scenarios and coarsening exponents.

{\it Phenomenology and model}.---Here we focus on deposition processes on a high symmetry substrate, where
atoms or molecules arrive ballistically %on the growing surface
and thermally diffuse until
they are incorporated into the crystal~\cite{RMP_Chaouqi}. 
A key process, which is the possible cause of the instability,
is the attachment to steps: if adatoms stick to ascending steps more efficiently than to
descending steps, an uphill current forms and the growing surface destabilizes,
forming a mound structure.
\comment{While time and length scales of the emerging structure can
be analyzed with a linear theory, the subsequent dynamics require a fully nonlinear treatment.}

Experimental results are available for many systems, specially metals, and different dynamical
scenarios have appeared. Cu(100)~\cite{Cu100} and Rh(111)~\cite{Rh111}
are examples of a coarsening process
which occurs during the whole experimental time scale. Instead, Pt(111)~\cite{Pt111} is
a prototype for coarsening which immediately stops or does not even start, while the height of mounds
goes on increasing. In some cases, slopes of mound tend to
constant values, corresponding to faceting. In other cases, observed slopes increase as well.
Finally, the symmetry of the pattern is related to the unstable orientation: (100) surfaces
produce a $4$-fold pattern, while (111) surfaces produce a $3$-fold pattern~\cite{111}.
The quantitative determination of the coarsening exponent, when appropriate,
is made extremely difficult by the short experimental time scale over which $L(t)$ is
measured and by the relatively smallness of the coarsening exponent, $n\lesssim \frac{1}{3}$.

Theoretical work has accompanied experiments with Kinetic Monte Carlo simulations~\cite{KMC}
and with continuum approaches~\cite{review_Paolo}, 
where the dynamics of the local height of the surface, $h(\bx,t)$,
is governed by a partial differential equation. Much effort has been devoted
to give robustness to the equations, while the equations themselves have been 
mainly studied numerically. Here below we propose
a more systematic and rigorous approach in the continuum limit,
starting from the well established class of equations:
\be
\label{eq.crescita}
\frac{\partial h(\mathbf{x},t)}{\partial t} = -\mathbf{\nabla}\cdot {[\mathbf{j}(\nabla h) +
\mathbf{\nabla}(\nabla^2 h)]} \equiv -\mathbf{\nabla}\cdot {\mathbf{J}_{tot}},
\ee
whose conservative form signals that, at the experimental conditions usually set for these deposition processes, overhangs and evaporation are negligible.
The total current, $\mathbf{J}_{tot}$, is composed of two terms: 
the nonequilibrium, slope dependent current, $\mathbf{j}(\nabla h)$,
describes the effects of
the Ehrlich-Schwoebel barrier, which hinders adatoms to stick to the lower step,
with $\mathbf{j}\simeq \nabla h$ for $|\nabla h| \ll 1 $~\cite{Villain}; 
the second term, $\mathbf{\nabla}(\nabla^2 h)$, 
is the Mullins term, describing termal detachment of adatoms from steps and
nonequilibrium effects~\cite{Rodi}.
In Eq.~(\ref{eq.crescita}), the white noise due to the flux can be omitted~\cite{Bray}.
We would like to highlight that we do not give explicit form to the slope dependent
$\mathbf{j}$: our analytical approach embraces any current $\mathbf{j}$.
We can even add that Eq.~(\ref{eq.crescita}) is also relevant 
for thermodynamically unstable surfaces~\cite{spinodal}.

{\it The method}.--- Equation (\ref{eq.crescita}) admits  the trivial solution $h\equiv 0$, corresponding to the flat profile. Setting
 $h=\delta~ \hbox{exp}(\omega t + i \mathbf{k} \cdot \mathbf{x})$ and linearizing in $\delta$  
yields  $\omega(k)=k^2-k^4$, $k=|\mathbf{k}|$,
%in FIG.~\ref{spectrum},
showing linear exponential growth, until the nonlinearity can no longer be neglected.%finché la parte non lineare prenderà il sopravvento e tutte le armoniche si mescoleranno....

The amplitude growth of mounds is thus fast, while wavelength rearrangement (if any) is  slow and  follows a diffusion process, as seen below. At short time scale the structure assumes a periodic pattern defined by two basis wave vectors $\mathbf{q}_1$,$\mathbf{q}_2$, or, similarly, by the two phases $\varphi_{i}=\mathbf{q}_{i} \cdot \mathbf{x}$.
%In order to study perturbations of these states (and so variations of the typical length scale of the system), we use a multiple scale technique.
The slow evolution of the phase makes it legitimate to introduce, besides the fast variable $\mathbf{x}$, a slow dependence on time and space, $T$ and $\mathbf{X}$, so that the local wave vector $\mathbf{q}=\mathbf{q}(T,\mathbf{X})$.
%and the phase variables will evolve, according to two coupled linear diffusion equations. It is a perturbative method, refined by the definition not only of series for the unknow function, $h=h_0+\varepsilon h_1$ with $\varepsilon \ll 1$ as dimensionless parameter, but also fast and slow scales for each indipendent variables: $t$, $\mathbf{x}$ and
Owing to the diffusion character of the phase, we expect $T=\varepsilon^2 t$ and $\mathbf{X}=\varepsilon \mathbf{x}$, where $\varepsilon$ is a small parameter measuring the long wavelength modulation (small gradient) of the phase.
%Our choose for the scaling powers of
%$\varepsilon$ is motivated by the expectation of diffusion equations for the phases, but it is also possible to gain them \textit{a posteriori}, from a more %general reasoning.
%We also introduce slow variables for the phases:
It is convenient to introduce  slow phases, related to the fast one by $\psi_{i}=\varepsilon \varphi_{i}$, so that $\mathbf{q}_{i}=\nabla_{\mathbf{x}} \varphi_{i}=\nabla_{\mathbf{X}} \psi_{i}$.
In a multiscale spirit, from these definition together with $h=h_0+\varepsilon h_1+\dots$, the following expansions for the operators hold:
\bea
\partial_t &=& 
\varepsilon \left((\partial_T \psi_1) \partial_{\varphi_1} + (\partial_T \psi_2) \partial_{\varphi_2}\right) 
\label{nabla} \\
\nabla &=& \nabla_0 + \varepsilon \nabla_\mathbf{X} ,
\eea
with $\nabla_0 = \mathbf{q}_1 \partial_{\varphi_1} + \mathbf{q}_2 \partial_{\varphi_2}$ and $\nabla_{\mathbf{X}} = (\partial_X, \partial_Y)$.
\comment{
The Laplacian operator is then:
\begin{equation} \label{Laplac}
\nabla^2 = \nabla_0^2 + \varepsilon [\nabla_0 \nabla_\mathbf{X} + \nabla_\mathbf{X} \nabla_0]
\equiv \nabla_0^2 + \varepsilon \nabla_1^2 ,
\end{equation}
}
Finally, the current $\mathbf{j}$ reads:
\be
\mathbf{j}(\nabla h) = \mathbf{j}(\nabla_0 \th_0) + \varepsilon \mathcal{J} (\nabla_0 \th_1 + \nabla_\mathbf{X} \th_0) ,
\ee
where $\mathcal{J}$ is the Jacobian matrix. Reporting the above expansions into Eq.~(\ref{eq.crescita}) we obtain successively higher order contributions in powers of $\varepsilon$.
To order $\varepsilon^0$ we gain
\begin{equation} \label{ord.zero}
0 = \nabla_0 \cdot {[\mathbf{j}(\nabla_0 \th_0) + \nabla_0(\nabla_0^2 \th_0)]}
= \nabla_0 \cdot {(\mathbf{J}_0)_{tot}}
\equiv \mathcal{N}[\th_0] ,
\end{equation}
which is a nonlinear equation satisfied by the steady-state profile $\tilde h_0$. Since we are interested in growth on high symmetry substrates, the stronger condition $(\mathbf{J}_0)_{tot}=0$ is fulfilled.
Then, to first order we obtain: 
\be
\mathcal{L}[\th_1] = g(\th_0,\psi_1,\psi_2), 
\label{ord.uno}
\ee
where
\be
\mathcal{L}[\th_1] \equiv -\nabla_0 \cdot {[\mathcal{J}(\nabla_0 \th_1) + \nabla_0 (\nabla_0^2 \th_1)]} 
\ee
is the Fr\'{e}chet derivative of $\mathcal{N}$, and
\be
\begin{split}
g& \equiv (\partial_T \psi_1) \partial_{\varphi_1} \th_0 + (\partial_T \psi_2) \partial_{\varphi_2} \th_0
+ \nabla_0 \cdot [\mathcal{J}(\nabla_{\mathbf{X}} \th_0) \\
&+\nabla_{\mathbf{X}}(\nabla_0^2 \th_0) + \nabla_0(\nabla_1^2 \th_0)] .
\end{split}
\ee
In order to avoid secular terms, we use the Fredholm's alternative theorem \cite{zwillinger}, 
according to which Eq.~(\ref{ord.uno}) 
has solutions if and only if $\langle v,g \rangle=0$~\cite{scalarproduct}, where
${\cal L}^\dagger [v]=0$.

While the derivation of the phase equation can be achieved for any nonlinear equation,  
we focus here on Eqs.~(\ref{eq.crescita}) such that $\mathcal{L}$ is a self-adjoint operator,
$\mathcal{L}^{\dagger}=\mathcal{L}$. This condition is equivalent to saying that the
Jacobian ${\cal J}$ is symmetric, which, in fact, is not a restriction:
all surface currents $\mathbf{j}$ discussed in the literature provide a symmetric ${\cal J}$.
Thanks to the translational invariance of $\mathcal{N}$ with respect to the space variable, 
we easily obtain non trivial solutions of the form $v_i=\partial_{\varphi_i} h_0$, $i=1,2$ 
(Goldstone mode). This leads us to  two diffusion equations ($i=1,2$):
\be \label{eq.diff.gen}
\partial_T \psi_i = \frac{\partial \psi_{\alpha}}{\partial X_{\beta} \partial X_{\gamma}}
\tilde D_{\beta\gamma}^{i\alpha} ,~~~~~\alpha,\beta,\gamma=1,2
\ee
where the diffusion coefficients have the following expressions:
\be
\label{diffusion_coeff}
\tilde D_{\beta\gamma}^{1\alpha}=\left[\frac{\langle h_1,c^{\alpha}_{\beta\gamma}\rangle \langle h_2,h_2\rangle  - \langle h_2,c^{\alpha}_{\beta\gamma}\rangle \langle h_1,h_2\rangle }{\langle h_1,h_1\rangle \langle h_2,h_2\rangle -{\langle h_1,h_2\rangle }^2}\right]
\ee
and $\tilde D_{\beta\gamma}^{2\alpha}~ \substack{{1 \leftrightarrow 2} \\ \displaystyle =}~ 
\tilde D_{\beta\gamma}^{1\alpha}$.
Here above we have introduced the more compact notation $h_j =
\partial_{\varphi_j}\th_0= \partial_j \th_0$, and
\be
\begin{split}
-c^{\alpha}_{\beta\gamma} &= q_{\delta\nu}\partial_{\varphi_\delta}
\left[\mathcal{J}_{\nu\gamma}\frac{\partial h_0}{\partial q_{\alpha\beta}}\right]
+ 2 q_{j\gamma}q_{l\beta}\partial_{\alpha} \partial_l \partial_j h_0 \\
&+ 3 \nabla_0^2 q_{\nu\beta} \partial_{\nu} \frac{\partial h_0}{\partial q_{\alpha\gamma}}
+ \delta_{\beta\gamma} \nabla_0^2 \partial_{\alpha} h_0 ,
\label{c_generali}
\end{split}
\ee
where the use of the zeroth order relation $\mathbf{j}(\nabla_0 \th_0) = - \nabla_0(\nabla_0^2 \th_0)$
has allowed to replace the current with derivatives of the steady profile.

We conclude this part stressing that our $\epsilon-$expansion might continue at higher order,
leading to nonlinear terms in Eqs.~(\ref{eq.diff.gen}). We refer the reader to the discussion in
\cite{1d}, Sect. VI\.C. 

\textit{Pattern symmetries and phase diffusion equations}---For definiteness we focus  
in the following on square (4-fold) and hexagonal (6-fold) patterns.
In the square case, we set $\mathbf{q}_1=q(1,0)$ and $\mathbf{q}_2=q(0,1)$ and exploiting parity and $4$-fold symmetry for $\th_0$, physical properties are equivalent upon the changes $\varphi_1 \leftrightarrow \varphi_2$. 
In the hexagonal case, we have $\mathbf{q}_1=q(1/2,\sqrt{3}/2)$ and $\mathbf{q}_2=q(1/2,-\sqrt{3}/2)$ 
and parity and $6$-fold symmetry imply invariance under the changes  $(\varphi_1 \leftrightarrow \varphi_2)$
 and  $(\varphi_1 \rightarrow -\varphi_2$, $\varphi_2 \rightarrow \varphi_1+\varphi_2)$.
The interesting result is that, for both symmetries, Eqs.~(\ref{eq.diff.gen}) assume 
a much simpler form \cite{diff.coeff}:
\bse \label{eq.diff.fin}
\be
\partial_T \psi_1 = (\psi_1)_{11} D_{11} + (\psi_1)_{22} D_{22} + (\psi_2)_{12} D_{12} 
\ee
\vskip -0.8cm
\be
\partial_T \psi_2 = (\psi_1)_{12} D_{12} + (\psi_2)_{11} D_{22} + (\psi_2)_{22} D_{11} ,
\ee
\ese
with only three non vanishing diffusion coefficients. 
\comment{
since $D_{11}^{11} = D_{22}^{22} \equiv D_{11}$, $D_{22}^{11} = D_{11}^{22} \equiv D_{22}$, $D_{12}^{12} = D_{12}^{21} \equiv D_{12}$ and the others ones are zero.
}
The hexagonal pattern enjoys an additional property:
\be \label{famosa}
D_{11}-D_{22}-D_{12}=0,
\ee
leaving us with only two independent diffusion coefficients. 

Phase instability means wavelength rearrangement, leading to coarsening. To investigate stability of the pattern we set
$\psi_{1,2}(\mathbf{X},T)=\psi_{1,2}^{(0)} \exp{(\Omega T)}\exp{(i\mathbf{K}\cdot\mathbf{X})}$,
where instability is signaled by at least one positive eigenvalue $\Omega_{1,2}(\mathbf{K})$. 
%so the spectrum $\Omega=\Omega(K)$ is invariant under rotations, while in the square case this is not valid and $\Omega=\Omega(K,\theta)$. This constitutes a %formal analogy with the elasticity theory in 2D \cite{Walgraef} and it also proves that the hexagonal symmetry case is closed to the isotropic one %\cite{golub00}.
In the hexagonal case we straightforwardly find from Eqs.~(\ref{eq.diff.fin}) that:
\begin{equation}
\Omega_1=-D_{22}K^2 \, , ~~ \Omega_2=-D_{11}K^2.
\end{equation}
Since $D_{ij}$ are expressed in terms of integrals involving the steady periodic pattern $h_0$ (see Eq.(\ref{diffusion_coeff})), stability can be linked to the property of steady-state solutions only. 
It is a simple matter to show that $D_{22}$ is positive, thus $\Omega_1<0$. The analysis of $D_{11}$ is much more involved. After several manipulations we find:
\begin{equation} \label{D.esag}
D_{11}=\frac{4q^{7/4}}{\langle h_1^2\rangle } \partial_q (q^{5/4}\langle h_{12}^2\rangle ) ~\equiv~ f(q) \mathcal{A}'(q).
\end{equation}
The sign of $D_{11}$ (and thus of $\Omega_2$) is given by the sign of the slope of the function
%With the last passage we bring out the possibility to relate the sign of the coefficient to a total derivative of a function
$\mathcal{A}=(q^{5/4}\langle h_{12}^2\rangle )$ with respect to $q$. In the 1D models studied in \cite{1d} this function was found to be the amplitude %usare il termine amplitue è pericoloso, vedi cuccoli...
of the steady state solutions. The present study reveals a new class of dynamics where the relevant quantity is $\mathcal{A}(q)$, having a more abstract meaning than just the amplitude itself. At present a simple physical interpretation of this function is missing.

For square symmetry, the spectrum is anisotropic, depending on the angle $\theta$ 
between $\mathbf{K}$ and the $X$ axis. However, symmetry imposes that the growth rate may be 
maximal along two directions: (i) $\theta=0$, for which $\Omega_1^0 (K) = -D_{22} K^2$ 
and $\Omega_2^0 (K) = -D_{11} K^2$, and (ii) $\theta=\pi/4$, for which  $\Omega_1^{\pi/4} (K) 
= -(D_{11}+D_{22}-D_{12}) K^2/2$ and $\Omega_2^{\pi/4} (K) = -(D_{11}+D_{22}+D_{12}) K^2/2$.
Again, the stability  is dictated by the sign of diffusion coefficients:
we find that the eigenvalues 
$\Omega_1^{0,\pi/4}$ are negative, while the eigenvalues $\Omega_2^{0,\pi/4}$ have no fixed sign:
\bse
\label{autov.quad}
\be
D_{11} = \frac{1}{\langle h_1^2\rangle } [\partial_q (q^3 \langle h_{11}^2\rangle ) + q^3 \partial_q \langle h_{12}^2\rangle + q^2 \langle h_{12}^2\rangle  ] ,
\ee
\be
\begin{split}
D_{11}&+D_{22}+D_{12} = \frac{4}{\langle h_1^2\rangle } \left[ \frac{1}{2} q^3 \partial_q \langle h_{11}^2\rangle  + q^2 \langle h_{11}^2\rangle \right.\\
&+ \left. \frac{1}{2} q^3 \partial_q \langle h_{12}^2\rangle + 2q^2 \langle h_{12}^2\rangle \right].
\end{split}
\ee
\ese
\begin{figure}
\centering
\includegraphics[width=7cm]{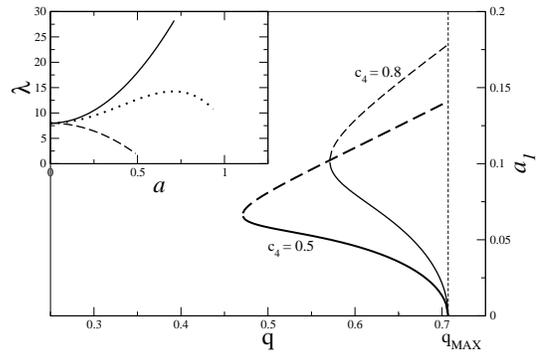}
\caption{The amplitude of stationary solutions as function of $q$, for square symmetry
($2q^2_{\hbox{\tiny MAX}}=k^2_{\hbox{\tiny MAX}}=1$): 
the full line indicates the branch along which the system will coarsen; along the dashed line there is no coarsening. Branches marked by the thinnest line correspond to a larger value of $c_4$. 
%{\color{red} may be give values of $C_i$ on the graph itself.}. 
Inset: the three possible scenarios for the growth model in 2D found in the limit $k \to k_{\hbox{\tiny MAX}}$, in which the sign of the diffusion coefficients that can give instability is linked to the behaviour of the amplitude $a$ with respect to the wavelength $\lambda$. 
We have coarsening (full line), no coarsening (dashed line), and interrupted coarsening (dotted line). }
\label{collage}
\end{figure}

{\it Coarsening scenarios and coarsening exponents}.---Even if an exact determination of diffusion
coefficients can only be done numerically, there are two crucial limits allowing analytical
treatment: the small amplitude limit $(k\to k_{\hbox{\tiny MAX}} =1)$ and the large
wavelenght/large amplitude limit. The former provides the proof of the existence of different
coarsening scenarios, the latter provides the explicit values of coarsening exponents.

For the small amplitude limit, we have considered an isotropic current of the form 
$\mathbf{j}(\mathbf{m},c_2,c_4)=
\mathbf{m}(1+c_2 \mathbf{m}^2+c_4 \mathbf{m}^4)$ and we have found~\cite{SM} steady solutions
and diffusion coefficients for both square and hexagonal symmetries.
In both cases, the phase stability is controlled by the amplitude $a_1(q)$ of the steady state:
we have instability, i.e. coarsening, if $a'_1(q) < 0$ (see Fig.~1). The sign of this derivative is 
determined by the coefficients $c_{2}$ and $c_4$: perpetual coarsening occurs if $c_2<0, c_4\le 0$ (or $c_2>0, c_4<0$), no
coarsening if $c_2>0, c_4>0$, coarsening up to a critical wavelenght (interrupted coarsening) if
$c_2<0, c_4>0$. The basic three scenarios are illustrated in the inset of Fig.~1. 

Perpetual coarsening is determined by a phase instability occurring for any $q$.
In this case we can extract the coarsening exponent $n$, $L(t)\approx t^n$, through
the following dimensional law:
\be \label{dimensional}
|D(q)| \approx \frac{L^2}{t},
\ee
with $q=2\pi/L$. The application of Eq.~(\ref{dimensional})
has always given values in agreement with known results,
both in 1D~\cite{1d} and in 2D~\cite{2d}. 

We insist on highlighting that the form of $\mathbf{j}$ is arbitrary and that here we shall only distinguish between two broad classes of systems: 
those which exhibit slope selection and those which do not.
In the first case the current $\mathbf{j}(\mathbf{m})$ has zeros for finite values of the slope  
$\mathbf{m}^*$ (as, for example, with $\mathbf{j}=\mathbf{m}(1-\mathbf{m}^2)$), so that pyramids will grow in size but with a slope tending to the constant value $\mathbf{m}^*$, leading to faceting. 
This implies that $\mathbf{m}^*=(\partial_x h_0,\partial_y h_0)$ is constant, except along domain walls,
that have a finite but small thickness. We first consider the square symmetry, so, for example:
\be
\langle h_{11}^2 \rangle= \frac{1}{(2\pi)^2} \int_{0}^{\lambda}dx\int_{0}^{\lambda}dy \frac{1}{q^2} 
\left( \frac{\partial^2 h_0}{\partial x^2}\right)^2 = c_{11} \lambda^3 + o(\lambda^3)
\nonumber
\ee
where $c_{11}$ is a positive constant.
With the same reasoning, at the leading order in $\lambda$ we find $\langle h_{12}^2 \rangle = c_{12} \lambda^3$ and  
$\langle h_{1,2}^2 \rangle = c_{\varphi} \lambda^2$.
Therefore, we get $\displaystyle D_{11} = -2q c_{12}/c_{\varphi}$,
that is negative, signaling instability.
For the other coefficient we have $(D_{11}+D_{22}+D_{12}) =2q (c_{12}-c_{11})/c_{\varphi}$,  
with $c_{11}>c_{12}$ for phase instability.
Plugging the results for the $D_{ij}$'s into (\ref{dimensional}) we finally obtain:
\be
L \sim t^{1/3}.
\label{n1_3}
\ee
A similar analysis for hexagons allows us to determine (\ref{autov.quad}) that has exactly the same behaviour with $\lambda$ as for the square symmetry, meaning that  coarsening exponent  $n=1/3$
also holds for 6-fold symmetry.

We have also analyzed the coarsening exponents for models without slope selection in the steady pattern: when $L$ increases, the slope increases as well. Assuming that the mound profile changes only along one direction while remains constant along the perpendicular one, we can find the asymptotic profile.
%solve the stationarity condition in only 1D. 
Thus, for a current that behaves asymptotically (large slope)  
as $\mathbf{j}(\mathbf{m})\simeq 1/|\mathbf{m}|^{\beta}$, $\beta>1$, we have $\displaystyle 1/|\mathbf{m}|^{\beta}=- \partial^2 m/\partial x^2$.
By using its solution,
% analogy with the Newton law, 
we have finally obtained the result:
\begin{equation}
L \sim t^{1/4}
\label{n1_4}
\end{equation}
for both symmetries and for any values of $\beta$, as in the 1D case \cite{1d}. 

{\it Conclusions and discussion}---Time and length scales of the emerging structure have been 
investigated within the phase diffusion notion, which has allowed to relate the growth dynamics 
to the properties of stationary solutions. The sign of the coefficients that dictate instability 
is related to some  abstract function (see Eq.~(\ref{D.esag})) that depends on steady-state solutions. 
In the limit of small amplitude the abstract function coincides with the amplitude of the pattern. 
In this case three kinds of scenarios, namely perpetual coarsening, no coarsening, and interrupted 
coarsening, are possible. This is the first analytical evidence of such dynamical
scenarios in 2D growth models.

As for the coarsening exponents, they deserve a special discussion. In fact, while the result
(\ref{n1_3}) for a faceted 6-fold pattern and the result (\ref{n1_4}) for an unfaceted pattern
agree with numerics and scaling considerations~\cite{Golubovic1997,golub00},
the exponent $n=1/3$ for a faceted square pattern seems to be
in contradiction with some numerics~\cite{golub00,siegert98}, where a slower coarsening $n=1/4$
is reported. 
\comment{
\textit{These authors argued  that in the square case the coarsening dynamics is slaved to 
defects annihilations occurring in $4$-fold pattern. Our analysis does not require defects generation. 
Actually, we have shown rigorously that the square  pattern is linearly unstable with respect to 
phase fluctuations. This contrasts with a previous suggestion~\cite{siegert98} according to which 
the pattern is metastable and therefore it would require defect generation to overcome 
the metastability barrier. Our exact results support clearly the idea that coarsening can take place
via pyramid coalescence without the assistance of defect generation. 
This statement is also consistent with some more recent numerical results obtained for specific 
expressions of the current  $\mathbf{j}$ \cite{GolubReview}. 
However, defect occurrence in some previous numerical simulations is beyond any 
doubt and it is found to slow down coarsening. Our study, which shows 
linear instability of the pattern,} 
}
These authors argue that in the square case the coarsening dynamics 
may be slaved to the appearance of {\it roof-tops}~\cite{siegert98},
a type of domain wall which is not present in a regular lattice of square pyramids,
where only the so called {\it pyramid edges} appear~\cite{note_dw}.
This type of dislocations, according to~\cite{golub00,siegert98}, 
plays a major role in their numerical simulations, leading to a slowing of
coarsening: $n=1/4$ instead of $n=1/3$.
However, more recent results~\cite{GolubReview} show that such topological defects
may or may not appear, depending on the explicit form of the slope dependent current.
Therefore, from the point of view of numerical simulations, the faceted square case
does not give a unique picture.
We should also add that in \cite{golub00,siegert98} the square pattern was
pointed out as metastable
and therefore defect generation was required to overcome the metastability barrier, 
but we have shown exactly that the square  pattern is linearly unstable with respect to 
phase fluctuations. 

A theory of coarsening in the presence of topological defects is beyond the reach 
of the present study. However, our results support the idea that coarsening can take 
place via pyramid coalescence, without the assistance of defect generation. 
Therefore, since defect occurrence was found to slow down coarsening, our analysis might suggest
that there always exists a faster channel for dynamics, 
i.e. without the intervention of topological defects. 
It would be an important future line of research to see whether the presence of defects or the lack 
thereof, is a fundamental ingredient for some specific equations, or rather does it depend on 
initial conditions, boundary conditions, system size, etc\dots

All these results have been established without evoking any specific form of the 
slope dependent current $\mathbf{j}(\mathbf{m})$, 
therefore pointing to the existence of a universality class. Furthermore, the results are 
independent of the considered symmetries: hexagons and squares. It would be an interesting task 
for future research to see whether or not this conclusion survives for the three other Bravais 
lattices (rectangular, rectangular centered and oblique).
Finally, we point out that unstable growth on a (111) surface produces a 3-fold pattern, which 
does not fall within Bravais lattices. It is an open question to understand how such pattern
can be analyzed within our approach.

A last word for experimental reults is appropriate. While the large variability
of experimental values for $n$~\cite{review_Paolo,111} prevents from a quantitative comparison,
our different dynamical scenarios for coarsening are well observed~\cite{Cu100,Rh111,Pt111}. 

{\it Acknowledgements}---CM thanks CNR for a Short-Term Mobility award from their International Exchange Program and CNES for financial support. Useful discussions with Matteo Nicoli are also acknowledged.


\begin{thebibliography}{99}
\bibitem{scalarproduct}
The scalar product is defined as $\displaystyle \langle f,g \rangle := \frac{1}{(2\pi)^2} \int_0^{2\pi} \int_0^{2\pi} d\varphi_1d\varphi_2 f^* g$.
\bibitem{diff.coeff}
Since the equality of mixed partials, it is preferable to define new diffusion coefficients
\be \label{convenz.coeff}
D^{i\alpha}_{\beta\gamma} = \left\{
\begin{array}{cc}
\tilde D^{i\alpha}_{\beta\gamma} & \beta = \gamma \\
\tilde D^{i\alpha}_{\beta\gamma} + \tilde D^{i\alpha}_{\gamma\beta} & \beta \neq \gamma
\end{array}
\right.
\ee
and to assume $\beta \leq \gamma$ in the diffusion equations (\ref{eq.diff.gen}).
\end{thebibliography}
\end{document}